\documentclass[twocolumn,showpacs,preprintnumbers,amsmath,amssymb]{revtex4}

\usepackage{graphicx}
\usepackage{dcolumn}
\usepackage{bm}
\input epsf

\begin{document}

\title{STEP POTENTIALS FOR DARK ENERGY}

\author{VICTOR H. CARDENAS}
\email{victor.cardenas@uv.cl} \affiliation{Departamento de F\'isica
y Astronom\'ia, Universidad de Valpara\'iso, Gran Breta\~na 1111,
Valpara\'iso, Chile}
\author{MARCO RIVERA}
\email{marivera@userena.cl} \affiliation{Departamento de F\'isica,
Facultad de Ciencias, Universidad de la Serena, Av. Cisternas 1200,
La Serena, Chile}

\begin{abstract}
We consider a reconstructing scheme using observational data from
SNIa, BAO and CMB, based on a model of dark unification using a
single non-minimally coupled scalar field. We investigate through a
reconstruction program, the main features the current observational
data imposes to the scalar field potential. We found that the form
suggested by observations implies a step feature in the potential,
where the kinetic and potential energy becomes of the same order of
magnitude. 
\end{abstract}


\maketitle

\section{Introduction}

Dark energy is the name of the unknown component responsible for the
current accelerated expansion of the universe \cite{dereview}. In
its simpler form, this can be described by a fluid with constant
equation of state parameter $w=-1$, corresponding to a cosmological
constant, leading to the successful $\Lambda$CDM model, the simplest
model that fits a varied set of observational data.

This model posses a high dependence to initial conditions that makes
it unnatural in many ways. For example, the current value for
$\Omega_{\Lambda}$ and $\Omega_M$ are of the same order of
magnitude, a fact highly improbable, because the dark matter
contribution decreases with $a^{-3}$, with $a(t)$ the scale factor,
meanwhile the cosmological constant contribution have had the same
value always. This problem in particular is known as the cosmic
coincidence problem.

In this context, the most natural way to understand the acceleration
of the universe, is to assume the existence of a dynamical
cosmological constant, or a theoretical model with a dynamical
equation of state parameter ($p/\rho=w(z)$). The source of this
dynamical \textit{dark energy} could be both, a new field component
filling the universe, as a quintessence scalar field \cite{quinta},
or it can be produced by modifying gravity \cite{modgrav}.

In \cite{Shafieloo:2009ti} the authors suggested that the current
observational data favor a scenario in which the acceleration of the
expansion has past a maximum value and is now decelerating. The key
point in deriving this conclusion is the use of the
Chevalier-Polarski-Linder (CPL) parametrization
\cite{Chevallier:2000qy}, \cite{Linder:2002et} for a dynamical
equation of state parameter
\begin{equation}\label{cpl}
 w(a)=w_0+(1-a)w_1,
\end{equation}
where $w_0$ and $w_1$ are constants to be fixed by observations and
$a$ is the scale factor. An update analysis performed using recent
SNIa data was informed in \cite{Li:2010da} where similar conclusions
were derived. Both analysis assume a flat universe.

In \cite{Cardenas:2011a}, we study the previous model, using a new
set of data, the Union 2 data set \cite{Union2}, and also
considering the possibility to incorporate the curvature parameter
$\Omega_k$, as a new free parameter in the analysis. We find that,
the three observational test; SNIa, BAO and CMB, all can be
accommodated in the same trend, assuming a very small value for the
curvature, $\Omega_k \simeq -0.08$. The best fit values suggest that
the acceleration of the universe has already reached its maximum,
and is currently moving towards a decelerating phase.

Using a scalar field to model dark energy, it is possible to
reconstruct the scalar field potential from observations. There are
many approaches to do this \cite{Sahni:2006pa}. Considering models
of a single non-minimally coupled scalar field, the rapid variations
of the equation of state parameter at low redshift can not be
described.

%
The quest for the unification of inflation, dark matter, and dark
energy, in different combinations, by a single field, has been
studied in
Refs.~\cite{Peebles:1998qn,Lidsey:2001nj,Padmanabhan:2002sh,Scherrer:2004au,Matos:2005yt,Arbey:2006it,Cardenas:2006py,Cardenas:2007xh,Panotopoulos:2007ri,Liddle:2006qz,Liddle:2008bm,Bose:2009kc,BasteroGil:2009eb,DeSantiago:2011qb}.
The main motivation behind all proposals is that we do not yet
understand the nature of the components responsible for the three
phenomena, but we do know that their special properties are beyond
the realm of the ordinary matter described by the Standard Model of
Particle Physics.

An extreme, most economical, possibility is that all three phenomena
can be explained by the existence of one single field. As was first
put forward in Ref.~\cite{Liddle:2006qz,Liddle:2008bm}, the simplest
option at hand is a scalar field $\phi$ with a potential of the form
$V(\phi) = V_0 + (1/2) m^2 \phi^2$. The energy scale $V_0$ is to be
set at the tiny value of the observed cosmological constant
considered in the concordance $\Lambda$CDM model, and the mass scale
of the field, $m \simeq 10^{-6} m_{\rm pl}$, is determined by the
amplitude of primordial perturbations generated during inflation.


It is the purpose of this paper to explore the consequences of this
trend, suggested from the observations, and look for special
features through a characterization of a unification model based on
a scalar field, that allows these transitions at low redshift. The
reconstruction program uses as an intermediate phase a certain
parametrization of $w(z)$, which after its test using the data, is
slightly altered looking for improvements in the fit. Here we use a
$\chi^2$ test considering the AIC and BIC criteria that controls the
number of parameters to be used.

\section{$w(z)$ from the observations}

The problem of extracting information of $w(z)$ from observations
can be understand in the following way. Because most of the
measurements give us information of the Hubble function $H(z)$ or
also the luminosity (or angular diameter) distance $d_L(z)$, we are
forced to use
\begin{equation}\label{wzhz}
   w(z)=-\frac13\frac{2(1+z)HH'-3H^2}{ H_0^2(1+z)^3\Omega_m-H^2},
\end{equation}
in the case of data from $H(z)$, and we are forced to use
\begin{equation}\label{wzdlz}
   w(z)=-\frac{1}{3}\left(\frac{2(1+z)D''+3D'}{D'-(D')^3\Omega_m (1+z)^3}\right),
\end{equation}
in the case of the luminosity distance, where $D\equiv H_0
d_L(z)/c$. Notice that the precision in values for $\Omega_m$ and
$\Omega_k$ are crucial in this reconstruction procedure.

The limitations of this process, first identified in
\cite{Maor:2000jy}, are related to the dependence of the function
$w(z)$ of first and second derivatives of the $H(z)$ and $D(z)$
functions respectively. In order to tackle this problem, we can try
to model the $w(z)$ shape, by using a proper parametrization. The
most used is the already mentioned CPL \cite{Chevallier:2000qy},
\cite{Linder:2002et}, but there are many others designed to specific
goals. For example, to describe a fast transition at redshift $z_t$
we can use
  \begin{equation*}
    w(z)=w_i + \frac{w_f - w_i}{1+\exp(\frac{z-z_t}{\delta})}
  \end{equation*}
or if we are interested in oscillatory behaviour we can use
\begin{equation*}
    w(z)=w_0 + w_1\cos\left(A\log \frac{1+z_c}{1+z}\right).
\end{equation*}
However, there is a concern regarding the number of parameters used
to parameterize $w(z)$. It is clear that increasing the number of
parameters \cite{Liddle:2004nh}, is easiest to improve the fit with
observations, but is not clear if in the meantime we are adjusting
noise, instead of the truly physical relation.

\section{The approach}

In this section we describe the method to reconstruct the scalar
field potential through the use of a iterative program improving the
fit using two parameterizations for the equation of state parameter
$w(z)$. In this section we assume that $\Omega_k=0$.

The comoving distance from the observer to redshift $z$ is given by
\begin{equation}\label{comdistance}
r(z) =  \frac{c}{H_0} \int_0^z \frac{dz'}{E(z')},
\end{equation}
where
\begin{eqnarray}\label{edez}
  E^2(z) & = & \Omega_m (1+z)^3+\Omega_{de}f(z), \\
  f(z) & = & \exp \left\{ 3 \int^z_0 \frac{1+w(z')}{1+z'} dz'
  \right\},\nonumber
\end{eqnarray}
and $\Omega_{de}=1-\Omega_m$. The SNIa data give the luminosity
distance $d_L(z)=(1+z)r(z)$. We fit the SNIa with the cosmological
model by minimizing the $\chi^2$ value defined by
\begin{equation}
\chi_{SNIa}^2=\sum_{i=1}^{557}\frac{[\mu(z_i)-\mu_{obs}(z_i)]^2}{\sigma_{\mu
i}^2},
\end{equation}
where  $\mu(z)\equiv 5\log_{10}[d_L(z)/\texttt{Mpc}]+25$ is the
theoretical value of the distance modulus, and $\mu_{obs}$ is the
corresponding observed one.

The BAO data considered in our analysis is the distance ratio
obtained at $z=0.20$ and $z=0.35$ from the joint analysis of the 2dF
Galaxy Redsihft Survey and SDSS data \cite{bao2}, that can be
expressed as
\begin{equation}
\frac{D_V(0.35)}{D_V(0.20)}=1.736\pm 0.065,
\end{equation}
with
\begin{equation}
D_V(z_{BAO})=\bigg[\frac{z_{BAO}}{H(z_{BAO})}\bigg(\int_0^{z_{BAO}}\frac{dz}{H(z)}\bigg)^2\bigg]^{1/3}.
\end{equation}
We fit the cosmological model minimizing the $\chi^2$ defined by
\begin{equation}
\chi_{BAO}^2=\frac{[D_V(0.35)/D_V(0.20)-1.736]^2}{0.065^2}.
\end{equation}
A result from the combination of SNIa and BAO is given by a joint
analysis finding the best fit parameters that minimize
$\chi_{SNIa}^2+\chi_{BAO}^2$.

In addition, we can incorporate to the analysis the CMB redshift
parameter \cite{cmb2}, which is the reduce distance at $z_{ls}=1090$
\cite{Wang:2007mza}
\begin{equation}
R=\sqrt{\Omega_{m} H_0^2}r(z_{ls}) = 1.71\pm 0.019.
\end{equation}
We also apply the $\chi^2$
\begin{equation}
\chi^2_{CMB}=\frac{[R-1.71]^2}{0.019^2}\;,
\end{equation}
to find out the result from CMB and the constraints from
SNIa+BAO+CMB are given by $\chi_{SNIa}^2+\chi_{BAO}^2+\chi_{CMB}^2$.

Assuming a form for $w(z)$ in terms of a certain number of
parameters, we perform a Bayesian analysis to obtain the best fit
values of all the free parameters in the model.

\section{The CPL case}

For example, using the CPL parametrization, $w=w_0+w_1 z/(1+z)$, the
function defined in (\ref{edez}) leads to
\begin{equation}
f(z)=(1+z)^{3(1+w_0+w_1)}\exp \left(-\frac{3w_1z}{1+z}\right).
\end{equation}
Using the Union 2 set \cite{Union2}, consisting in $557$ type Ia
supernovae, the analysis leads to the results shown in Table
\ref{tab:table01}.

\begin{figure}[h!]
\centering \leavevmode\epsfysize=4.5cm \epsfbox{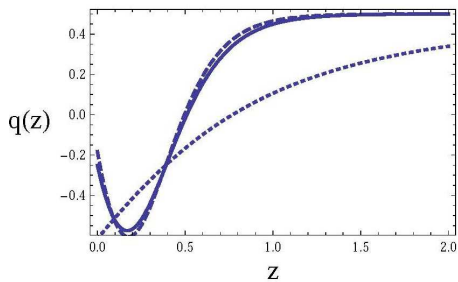}\\
\caption{ Using the Union 2 data set we plot the deceleration
parameter reconstructed using the best fit values for three cases:
only SNIa (continuos line), SNIa+BAO (dashed line) and SNIa+BAO+CMB
(dotted line)}\label{fig01}
\end{figure}

\begin{table}[h]
\caption{ The best fit values for the free parameters using the
Union 2 data set in the case of a flat universe model.}
{\begin{tabular}{@{}ccccc@{}} \toprule Data Set & $\chi^2_{min}$ & $\Omega_m$ & $w_0$ & $w_1$ \\
\colrule
SN & 541.43 & 0.4197 & -0.8632 & -5.490 \\
SN+BAO & 542.11 & 0.4281 & -0.7959 & -6.537  \\
SN+BAO+CMB & 543.91 & 0.2547 & -0.9979 & 0.190 \\ \botrule
\end{tabular}\label{tab:table01} }
\end{table}
A plot of the deceleration parameter using these numbers is shown in
Fig.\ref{fig01}.

\begin{figure}[h!]
\centering \leavevmode\epsfysize=5cm \epsfbox{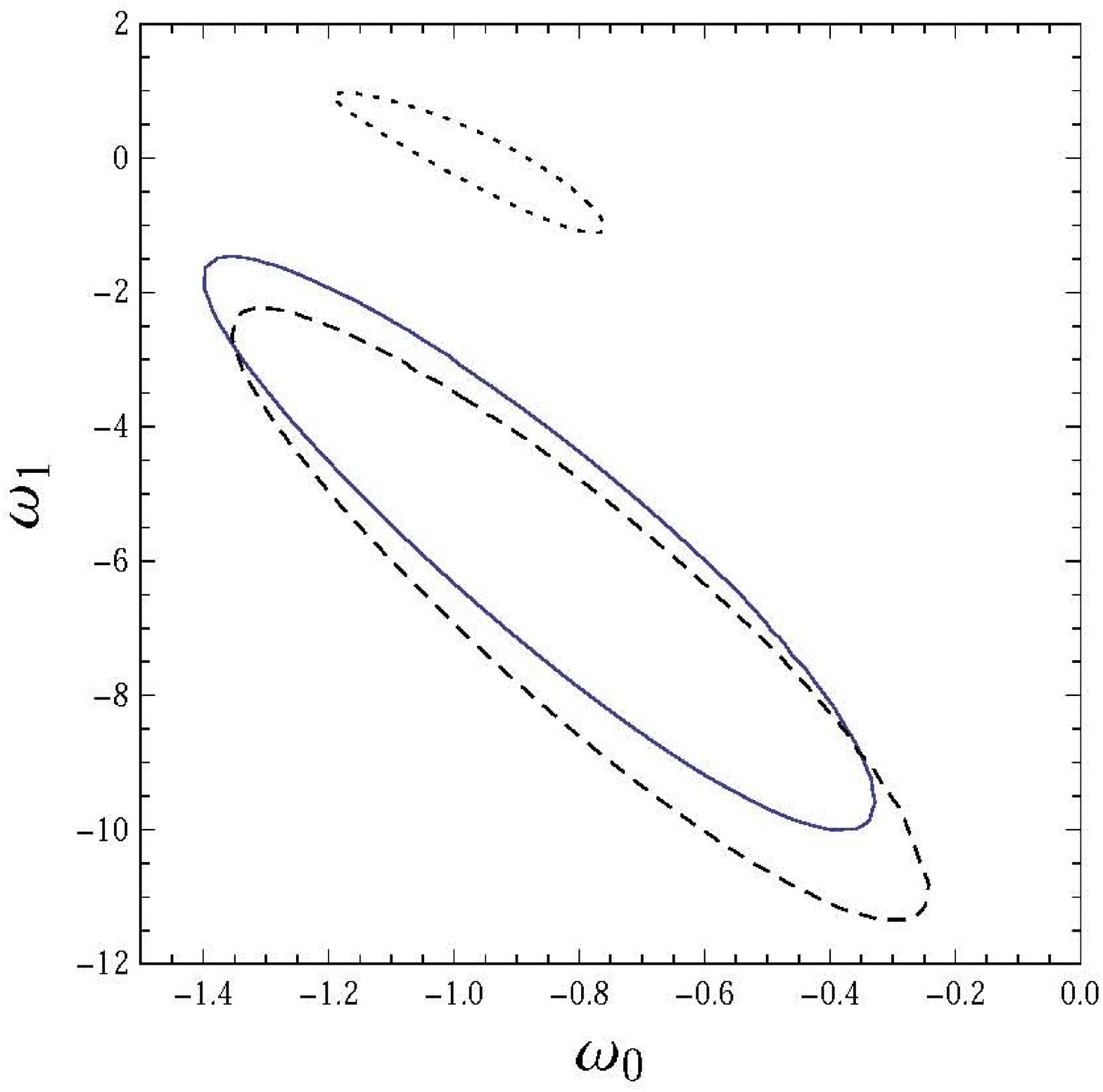} \caption{The
confidence limits for the case of Fig.{\ref{fig01}}}\label{fig02}
\end{figure}
The case SNIa alone and SNIa+BAO are almost identical, but the case
including CMB data does not show the rapid change at small redshift.
Using the best fit values for the parameters, give us the best fit
function $H(z)$ that we can use to reconstruct the scalar field
potential. Using a standard procedure \cite{Cardenas:2004ji} with
the equations for the flat case
\begin{eqnarray}\label{reconst}
    \phi'(z)^2 & = &\frac{1}{4 \pi G} \frac{H'}{(1+z)H}, \\
    V(z) & = &\frac{3}{8 \pi G}\left[ H^2 -
    \frac{(1+z)HH'}{3}\right],
\end{eqnarray}
we can plot directly the scalar field potential. In Fig.\ref{fig02}
we show the integration of these relations. As is expected, in the
cases SNIa and SNIa+BAO, we observe a rapid change in slope, a
``knee'' feature, that is not observed in the case including CMB.
\begin{figure}[h]
\centering \leavevmode\epsfysize=5cm \epsfbox{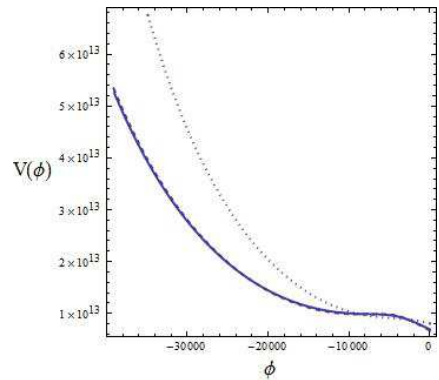}
\caption{Using the Union 2 data set we plot the scalar field
potential reconstructed using the best fit function $H(z)$ for the
three cases: only SNIa (continuos line),
 SNIa+BAO (dashed line) and SNIa+BAO+CMB (dotted line). The first
two overlap almost exactly.} \label{fig02}
\end{figure}
Then the sharp change  observed in the reconstructed deceleration
parameter $q(z)$ suggested by the data, is here visible in the
scalar field potential $V(\phi)$ as this knee feature
\cite{Wang:2009av}.

This is the anomalous effect mentioned first in
\cite{Shafieloo:2009ti} using the Constitution data set for SNIa,
and also in \cite{Li:2010da} and \cite{Cardenas:2011a}. A large
negative value for $w_1$ means that for small redshift the data
suggest a negative slope for $w(z)$. Because we are using the CPL
parametrization, this fact spoils the large $z$ behavior of the EoS
parameter, leading to a large negative values $w(z \rightarrow
\infty) \simeq -7$.

This is the kind of small redshift transitions in $w(z)$ that were
discussed first in \cite{Bassett:2002qu} and also in
\cite{Mortonson:2009qq}. In \cite{Bassett:2002qu} they use a form,
\begin{equation}\label{w2}
w(z)=w_0 + \frac{w_f - w_0}{1+\exp((z - z_t)/\Delta)},
\end{equation}
that captures the essence of a single transition at $z_t$ in a range
$\Delta$, from $w_0$ initially to a final value $w_f$ in the future.
In \cite{Mortonson:2009qq} the authors discussed the possibility of
a fast change in $w(z)$ at $z<0.02$ and its implication for a
standard scalar field model. They found that while a canonical
scalar field model can decrease the expansion rate at low redshift,
increasing the local expansion rate requires a non-canonical kinetic
term for the scalar field.

In this work we present an analysis of this problem using real data,
as opposite to the previous analysis, to constraint the form of the
equation of state parameter $w(z)$ and then through relations
(\ref{reconst}) to constraint the scalar field potential $V(\phi)$
in a unified dark matter dark energy model.

\section{Improving the fit}

As is evident from the previous section, the results show the
incompatibility of the CPL parametrization in describing the
variation of $w(z)$ with redshift, because the data suggest a very
large and negative value for $w_1$ which spoils the large $z$
behavior of $w(z)$.

In order to improve the fit, some authors have suggested the use of
a new parametrization \cite{Shafieloo:2009ti} for $w(z)$. We can use
for example, the fast single transition form (\ref{w2}), to improve
the fit. However, this parametrization has four parameters, two more
than the CPL. Adding a new parameter would be justifiable only if
the AIC or BIC (or a combination of these two) numbers indicate so
\cite{Liddle:2004nh}. In this context, if we do not have new
insight, we would like to keep the number of parameters fixed (in
this case two), until one of these numbers (AIC or BIC) indicated
something else.

Let us start with a first iteration of the process. The negative
value for $w_1$ obtained in the previous section indicate that
$w(z)$ has to change its slope recently, and as the CPL reproduce
perfectly, at large $z$ its value does not change very much. So, let
us use the following ansatz for $w(z)$:
\begin{equation}\label{w3}
w(z)=-\frac{w_0}{1+z}+\frac{w_1}{(1+z)^2}.
\end{equation}
This form has two parameters, like the CPL, so they are both
directly comparable with $\chi^2$ statistic, and also this has a
dust limit for $z\rightarrow \infty$. The main feature of this
parametrization is that allows the possibility to describe a change
of $w(z)$ at low $z$, without spoiling the large $z$ behavior. The
result of a bayesian analysis is displayed in Table
\ref{tab:table02}.
\begin{table}[h]
\caption{The best fit values for the free parameters of the
parametrization (\ref{w3}) using the Union 2 data set for SNIa, in a
flat universe model.} {\begin{tabular}{@{}ccccc@{}} \toprule
Data Set & $\chi^2_{min}$ & $\Omega_m$ & $w_0$ & $w_1$ \\
\colrule
N & 541.32 & 0.4121 & 8.7568 & 7.990 \\
SN+BAO & 542.11 & 0.4281 & -0.7959 & -6.537  \\
SN+BAO+CMB & 543.91 & 0.2547 & -0.9979 & 0.190 \\ \botrule
\end{tabular}\label{tab:table02}  }
\end{table}
Notice that, although the deceleration parameter obtained from SNIa
and SNIa+BAO still remains very close each other, meanwhile that
considering CMB behaves differently, the curve for the joint
analysis SNIa+BAO+CMB shows a change in slope close to $z=0$ that is
not observed using CPL.
\begin{figure}[h!]
\centering \leavevmode\epsfysize=5cm \epsfbox{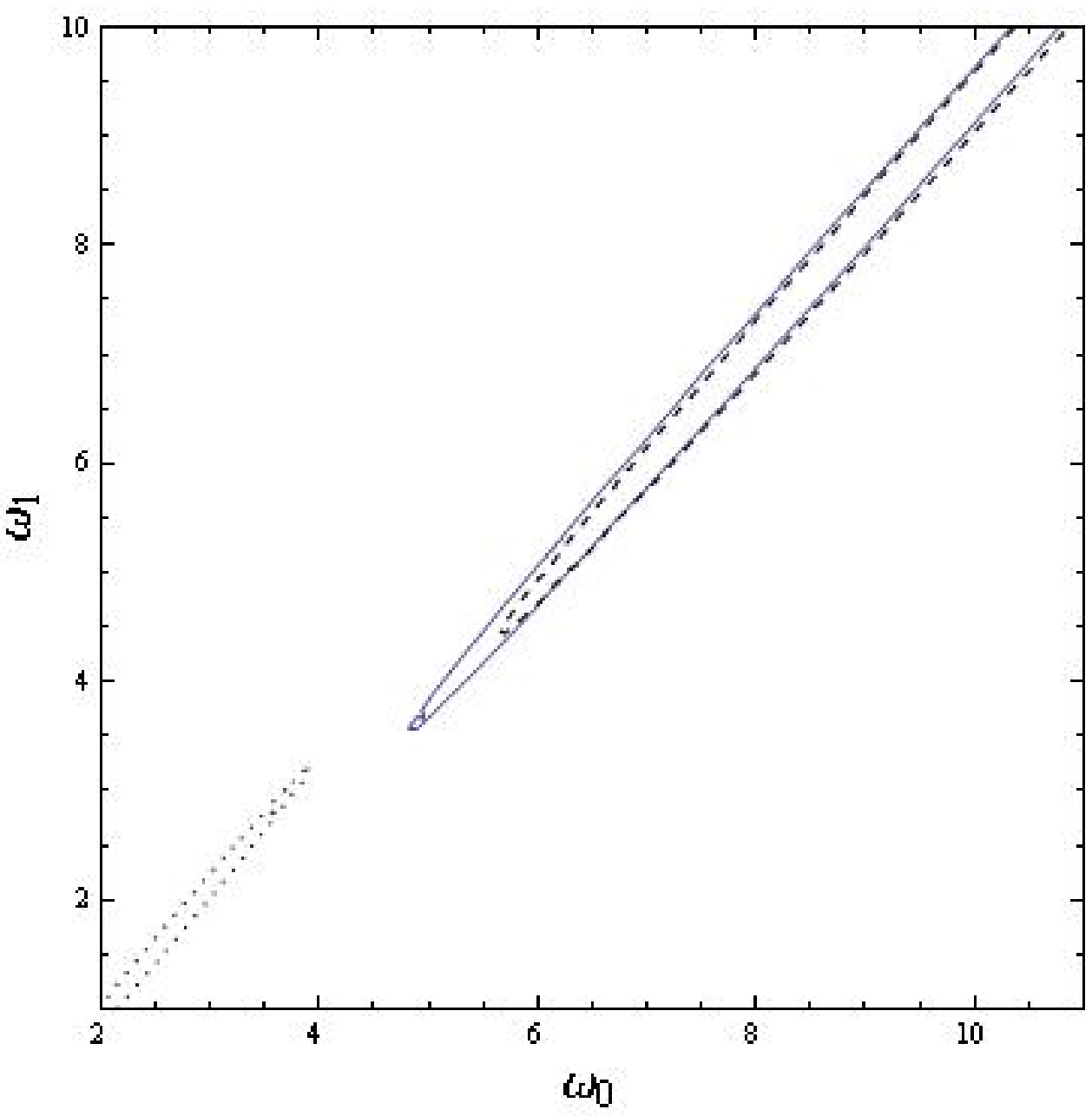}
\caption{The same as Fig.\ref{fig01}} \label{fig025}
\end{figure}

\begin{figure}[h!]
\centering \leavevmode\epsfysize=5cm \epsfbox{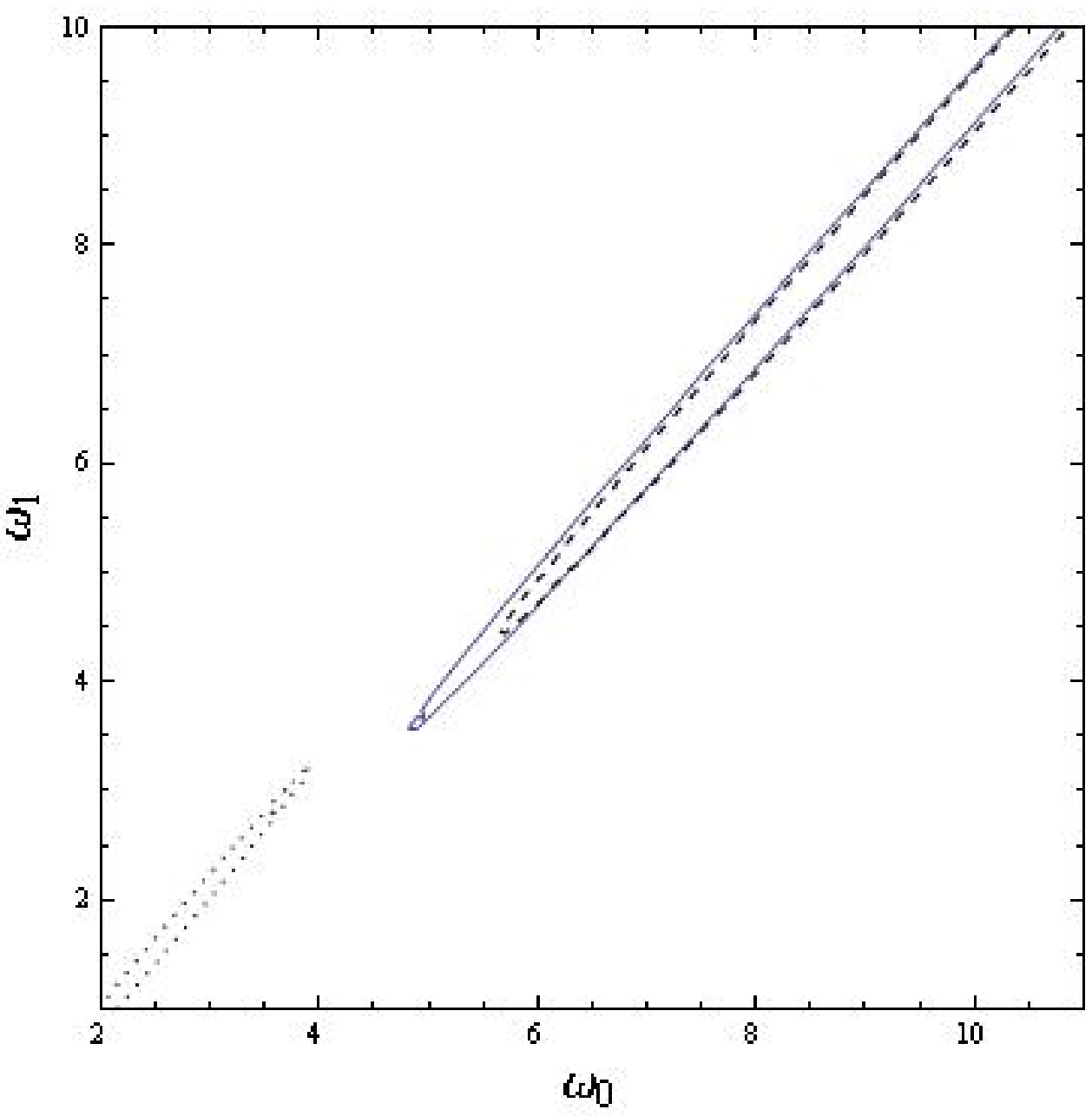}
\caption{The same as Fig.\ref{fig02}} \label{fig026}
\end{figure}

\begin{figure}[h!]
\centering \leavevmode\epsfysize=5cm \epsfbox{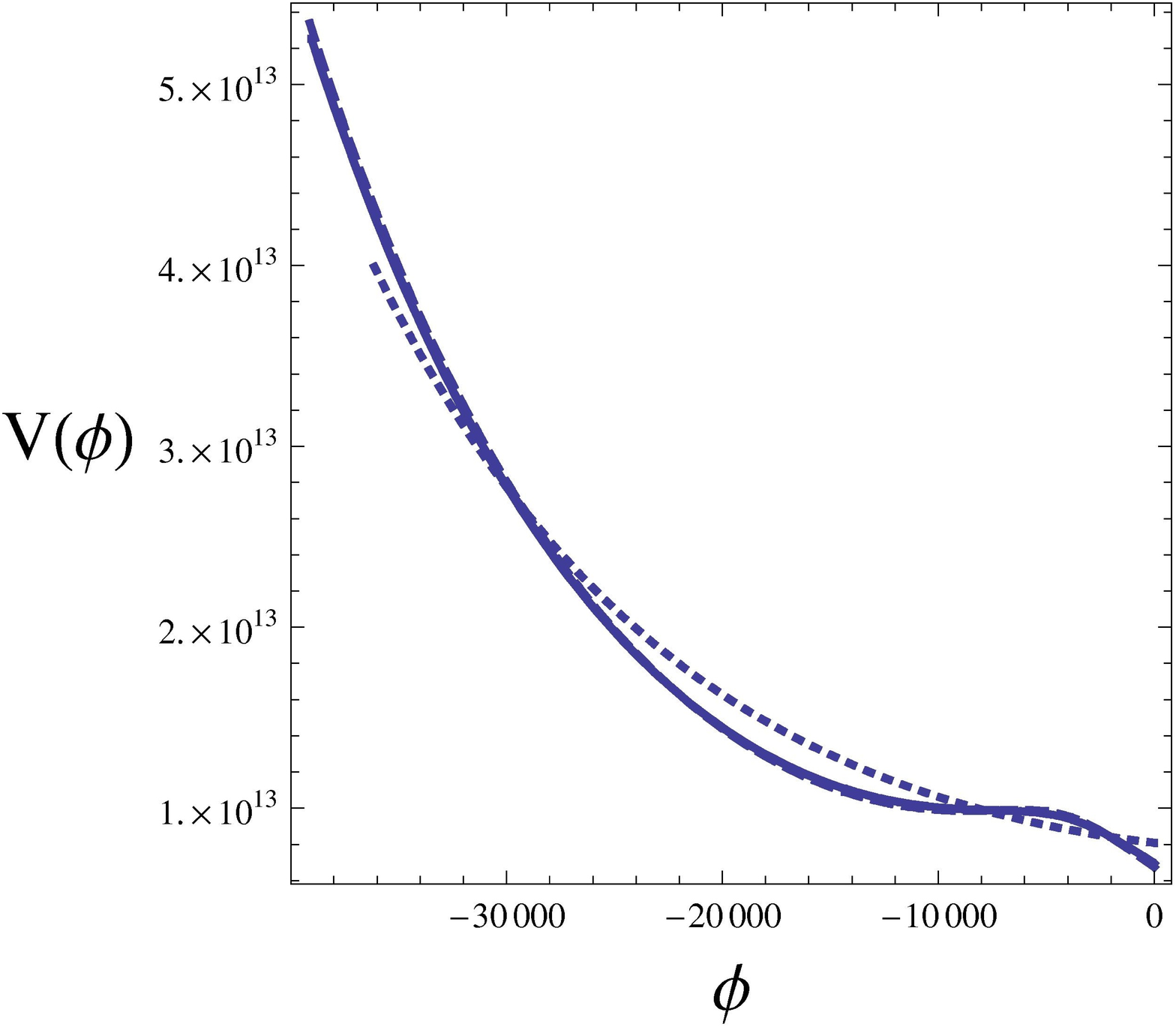}
\caption{\label{fig03} The same as Fig. \ref{fig02}. Notice the
improvement having all the curves overlap but still show a distinc
behavior once we consider the CMB data.}
\end{figure}

The second panel also shows this slight improvement, although still
remains the conflict between low and high redshift data. The
performance of this new parametrization is displayed in Table
\ref{tab:table03}.
\begin{table}[h]
\caption {A summary of the performance using CPL and the new
parametrization using only SNIa data. We display also the AIC and
BIC indexes to compare with the $\Lambda$CDM fit.}
{\begin{tabular}{@{}ccccc@{}} \toprule SNIa & $k$ & $-2\ln
\textsl{L} $ & $\triangle$AIC & $\triangle$BIC \\ \colrule
$\Lambda$CDM & 2 & 542.63 &  0     &   0   \\  
CPL &          3 & 541.43 &  0.80  &  -1.2  \\
New &          3 & 541.32 &  0.69  &  -1.31  \\ \botrule
\end{tabular}\label{tab:table03} }
\end{table}
Clearly, our new parametrization is an improvement respect to CPL.
The best fit parameters suggest a step like feature in the
potential, as we can see in Fig. \ref{fig03}. We can still look for
a better parametrization trying to accommodate all the three
observational probes in a single trend.
\section{Results}

In this paper we investigate the consequences of the low redshift
variations in the equation of state parameter, suggested by the
data, and their implications in a unified scalar field model for
dark matter and dark energy. This trend suggest a special feature in
the scalar field potential. The reconstruction program uses as an
intermediate phase a parametrization of $w(z)$, which after its test
using the data, is slightly altered looking for improvements in the
fit. Here we have used a $\chi^2$ test considering the AIC and BIC
criteria that controls the number of parameters to be used. We found
a better parametrization than the CPL, that properly describe the
low redshift behaviour, but although ameliorate the tension between
low and high redshift, it can not describe all the data together.

\section*{Acknowledgments}

The author wants to thanks Sergio del Campo for useful discussions.
VHC acknowledges financial support through DIPUV project No.
13/2009, and FONDECYT 1110230.


\end{document}